\documentclass{amsart}
\usepackage{bbm}
\usepackage[latin1]{inputenc}
\usepackage{verbatim}

\newtheorem{prop}{Proposition}
\newtheorem{lem}[prop]{Lemma}
\newtheorem{corol}[prop]{Corollary}
\newtheorem{theo}[prop]{Theorem}

\def\cal{\mathcal}

\def\R{{\mathbb R}}
\def\E{{\mathbb E}}

\def\P{{\mathbb P}}
\def\Var{\mathrm{Var}}

\def\eps{\varepsilon}
\def\etal{{\em et al.}}

\def\pT{{\widetilde{B}^\eps}}
\def\pB{\pT}
\def\pL{{\widetilde{L}^\eps}}

\def\wT{{B}}
\def\wB{\wT}
\def\wL{L}

\def\hT{{\widehat{B}}}

\newcommand\ind[1]{\mathbbm{1}_{\{#1\}}}

\title[Perturbation Analysis of a Variable $M/M/1$ Queue]{Perturbation Analysis of a Variable $\mathbf{M/M/1}$ Queue:\\ A Probabilistic Approach}

\address[Nelson Antunes, Christine Fricker, Philippe Robert]{INRIA-Rocquencourt,  RAP project, Domaine de Voluceau, 78153 Le Chesnay, France}

\address[Fabrice Guillemin]{France Telecom R\&D, CORE/CPN, 22300 Lannion, France}

\author[N.~Antunes]{Nelson Antunes}
\email{Nelson.Antunes@inria.fr}

\author[C.~Fricker]{Christine Fricker}
\email{Christine.Fricker@inria.fr}

\author[F.~Guillemin]{Fabrice Guillemin}
\email{Fabrice.Guillemin@francetelecom.com}

\author[Ph.~Robert]{Philippe Robert}
\email{Philippe.Robert@inria.fr}

\date{\today}


\begin{document}
\begin{abstract}
Motivated by  the problem  of the coexistence  on transmission links  of telecommunication
networks of  elastic and unresponsive traffic,  we study in  this paper the impact  on the
busy  period  of an  $M/M/1$  queue  of  a small  perturbation  in  the server  rate.  The
perturbation depends upon an independent  stationary process $(X(t))$ and is quantified by
means of  a parameter $\eps  \ll 1$. We  specifically compute the  two first terms  of the
power series  expansion in  $\eps$ of  the mean value  of the  busy period  duration. This
allows us  to study the  validity of the  Reduced Service Rate (RSR)  approximation, which
consists in comparing the perturbed $M/M/1$ queue with the $M/M/1$ queue where the
service rate is constant and equal to the mean value of the perturbation. For the first
term of  the expansion, the two systems are equivalent. For the second term,  the
situation  is  more  complex and  it  is shown  that  the  correlations of  the
environment process $(X(t))$ play a key role.
\end{abstract}
\maketitle

\bigskip

\hrule

\vspace{-1cm}

\tableofcontents

\vspace{-1cm}

\hrule

\bigskip

\section{Introduction}
We consider in this paper an $M/M/1$ queue with a time varying server rate. We
specifically assume that the server rate depends upon a random environment represented by
means of a process $(X(t))$, taking values in some (discrete or continuous) state space and
assumed to be stationary.  The study of this queueing system is motivated by the following
engineering problem: Consider a transmission link of a telecommunication network carrying
elastic traffic, able to adapt to the congestion level of the network, and a small
proportion of traffic, which is unresponsive to congestion. The problem addressed in this
paper is to derive quantitative results for  estimating the influence of unresponsive
traffic on elastic traffic. 

In real implementations, elastic traffic is controlled by the so-called transmission
control protocol (TCP), which has been  designed in order  to achieve a fair bandwidth
allocation among sufficiently long flows at bottleneck links. If we assume that the link
under consideration is the bottleneck, say, the access link to the network, then it is
reasonable to assume that bandwidth is distributed among the different competing elastic
flows according to the processor sharing discipline (see for instance Massoulié and
Roberts~\cite{Massoulie} and Delcoigne \etal~\cite{Proutiere}). Unresponsive traffic is then composed of small data transfers,
which are too short to adapt to the congestion level of the network. Throughout the paper,
it will be assumed that long flows arrive according to a Poisson process. 

With the above modeling assumptions, unresponsive traffic appears for elastic flows  as a
small perturbation of the available bandwidth. In addition, when there is no unresponsive
traffic, owing to the insensitivity property satisfied by the $M/G/1$ processor sharing
queue, the number of long flows is 
identical to the number of customers in an $M/M/1$ queue. Hence, in order to obtain a
global system able to describe the behavior of long flows in the presence of unresponsive
traffic, we study an $M/M/1$ queue with a time varying server rate, which depends upon
unresponsive traffic (for instance the number of small flows and their bit rate). The
problem is then to estimate the impact of unresponsive traffic on the performance of the
system. A classical issue  is in particular to investigate the validity of the so-called
reduced service rate (RSR) approximation, which states that everything happens as if the
server rate for long flows were reduced by the mean load of unresponsive traffic.  
RSR  approximation results (also called reduced  load equivalence)
have been shown to hold in a large number of queueing systems where some distributions are
heavy tailed see Agrawal \etal~\cite{Agrawal:01},  Jelenkovi\'c and
Mom\v{c}ilovi\'c~\cite{Predag} for example.

It is worth noting  that queueing systems with time varying server  rate have been studied
in the literature  in many different situations. In  Núñez-Queija and Boxma~\cite{Nunez1},
the authors  consider a queueing system  where priority is  given to some flows  driven by
Markov Modulated  Poisson Processes (MMPP) with  finite state spaces and  the low priority
flows share the  remaining server capacity according to  the processor sharing discipline.
By assuming that arrivals are Poisson and service times are exponentially distributed, the
authors solve  the system by means of  matrix analysis methods.  Similar  models have been
investigated in  Núñez-Queija~\cite{Nunez3,Nunez2} by still using
the quasi-birth and death process  associated  with  the  system  and  a  matrix
analysis.  In  this  setting,  the characteristics  of the  queue  at equilibrium  are
expressed in  terms  of the  spectral quantities of some matrices leading to potential
numerical applications.  More recently, priority queueing systems with fast dynamics, which
can be  described by means  of quasi-birth  and death processes,  have been studied  via a
perturbation analysis of a Markov chain by Altman \etal~\cite{Altman}. Boxma and
Kurkova~\cite{Boxma:07} studies the tail distributions of an $M/M/1$ queue with two service
rates. 

Getting qualitative  results for  queueing systems with  variable service rates 
to study, for example, the impact of the variability of the service rate on the
performances  of the system is rather difficult.  At the  intuitive level,  it is quite well known that  the
variability deteriorates them  but, rigorously speaking, only few  results are available.  The
main objective  of this paper is  to get  some insight on  these phenomena  by considering a
slightly perturbed system.  As it  will  be  seen, deriving such  an  expansion is  already
quite technical.  

In this paper,  it is assumed that the server  rate of the $M/M/1$ queue  is equal at time
$t$ to $\mu+\eps p(X(t))$ for some  function $p$, where $(X(t))$ is the process describing
the environment  affecting the  service rate. In  Fricker \etal~\cite{Fricker:10},  it has
been assumed  that the process $(X(t))$  is a diffusion process  and that $p(x)  = -x$. In
this paper,  the perturbation function  $p$ is quite  general and the  environment process
$(X(t))$   is  only  assumed   to  be   stationary  and   Markovian.   Moreover, we are
specifically interested in the power series expansion of mean busy period duration  in
$\eps$, which quantifies the magnitude of the  perturbation.  As far as the first order is 
concerned, the RSR  approximation is valid: The time-varying server  queue is identical to
an equivalent $M/M/1$  queue with a fixed  service rate equal to the  average service rate
$\mu+\eps  \E[p(X(0))]$. Combining the observation with the results obtained in Antunes
\etal~\cite{Antunes:02}, one can easily conclude, via a simple regenerative argument, that
the RSR holds for the mean number of customers in the queue.  The  analysis of  the second
order is  much more intricate; the 
correlations  of  the  process  $(X(t))$  play  a key  role  and,  consequently,  the  RSR
approximation is no more valid.

The organization of this paper is as follows: The model is described in
Section~\ref{model}. The first order term in the power series expansion of the mean busy
period duration is computed in Section~\ref{BPFsec}. The second order term is derived in
Section~\ref{BPSterm}. Applications of the results are discussed in
Section~\ref{applications}. Some basic elements of  the $M/M/1$ queue are recalled
in Appendix.

\section{Model}\label{model}
\subsection{Notation and Assumptions}
Throughout the paper $\wL(t)$ denotes the number of customers at time $t$ in an 
$M/M/1$ queue with arrival rate $\lambda$ and service rate $\mu$. The variable
$\wT$ denotes the duration of a busy period starting with one customer: Given
$\wL(0)=1$, 
\[
\wT=\inf\{s\geq 0: \wL(s)=0\}.
\]
It is assumed that the stability condition $\lambda < \mu$ holds. The invariant
distribution of  $(\wL(t))$ is geometrically distributed with parameter
$\rho=\lambda/\mu$. For $x\geq 1$, the variable $\wB_x$ denotes the duration of a busy
period starting with $x$ customers. By definition, $\wB_1\stackrel{\text{dist.}}{=} \wT$.
By convention, in the following when the variables $\wB$, $\wB_1$ and $\wB_1'$ are used in
the same expression, they are assumed to be independent with the same distribution as
$\wB$. This queue will be referred to as the standard queue denoted, for short, by S-Queue.  

For $\xi\geq  0$, ${\cal N}_\xi$  denotes a Poisson  process with intensity $\xi$  and for
$0\leq a<b$, ${\cal  N}_\xi([a,b])$ denotes the number of points of  this point process in
the interval $[a,b]$. In particular, ${\cal N}_\lambda$ will represent the arrival process
and ${\cal  N}_\mu$ the  process of the  services of  the S-Queue.  The  Poisson processes
${\cal N}_\lambda$ and ${\cal N}_\mu$ will be  assumed to be independent one of each other
and independent of  the modulating Markov process $(X(t))$. The  process $(\wL(t))$ can be
represented as the solution of the stochastic differential equation
\begin{align}
d\wL(t)\stackrel{\text{def.}}{=}\wL(t)-\wL(t-)&={\cal N}_\lambda([t,t+dt])
-\ind{\wL(t-)>0} {\cal N}\mu([t,t+dt])\notag \\
&=d\,{\cal N}_\lambda(t) -\ind{\wL(t-)>0} d\,{\cal N}_\mu(t),\label{Sdiff}
\end{align}
where $\wL(t-)$ is the left limit of $\wL(s)$ at $s\nearrow t$. For 
the representation of queueing Markov processes as solutions of stochastic differential
equations, see Robert~\cite{Robert:08}.
\medskip

\paragraph{\bf The perturbed queue}
In the following, we consider an $M/M/1$ queue with a service rate varying in time as a
function of some process $(X(t))$ taking values in some space, denoted by ${\cal S}$. We
assume that the process $(X(t))$ is an ergodic Markov process on ${\cal S}$. Typically,
the state space of the environment ${\cal S}$ is a finite or countable set when $(X(t))$
is a Markov Modulated Poisson Process (MMPP)  or ${\cal S}=\R$ in the case of a diffusion,
for instance an Ornstein-Uhlenbeck process (see Fricker \etal~\cite{Fricker:10}). The
invariant measure of the process $(X(t))$ is denoted by $\nu$. The Markovian notation
$\E_{x}(\cdot)$ will refer only to the initial state $x$ of the Markov process
$(X(t))$, therefore 
$\E_\nu(\cdot)$ will denote the expected value when the process $(X(t))$ is at equilibrium. 

The variable $\pL(t)$ denotes the number of customers at time $t$ in the  $M/M/1$ queue with
time-varying service rate. The process $(\pL(t),X(t))$ is a Markov process. The transitions
of the process $(\pL(t))$ are given by: If $\pL(t)=l$ and $X(t)=x$ at time~$t$, 
\[
l\to \begin{cases}
l+1 & \text{at rate } \lambda \\
l-1 & \phantom{at}''\phantom{rate} (\mu + \eps p(x))\ind{l>0} 
\end{cases}
\]
for some function $p(x)$ on the state space of the environment ${\cal S}$ and some small parameter $\eps \geq 0$.  When $p(x)>0$, this
implies that there is an additional capacity of service when compared to the S-Queue. On the
contrary, when $p(x)<0$, the server is with  a slower rate than in the S-Queue. 
The quantity $p^+(a)$ (respectively $p^-(a)$) is defined as
$\max(p(a), 0)$ (respectively $\max(0,-p(a))$). At time $t\geq 0$,  the additional capacity
is therefore $\eps p^+(X(t))$ and  $-\eps p^-(X(t))$ is the lost capacity. The
perturbation considered in this paper is regular, see Altman \etal~\cite{Altman}.

The variable $\pT$ is the duration of a busy period starting with one customer, that is,
given $\pL(0)=1$, 
\[
{\pT}=\inf\{s\geq 0: {\pL}(s)=0\}.
\]
For $x\geq 1$, the variable $\pB_x$ denotes the duration of a busy period starting with $x$ customers
($\pB_1\stackrel{\text{dist.}}{=} \pT$).
In the rest of this paper, we make the two following assumptions:
\begin{align}
&\mbox{the function $|p(x)|$ is bounded by a constant } M>0\tag{$\mathrm{H_1}$}\\
&\eps\sup\{|p(x)|: x\in{\cal S}\}<\mu .
\tag{$\mathrm{H_2}$}
\end{align}

The following proposition establishes that the length of the busy cycle is indeed
integrable. The rest of the paper is devoted to the expansion of its expected value with
respect to $\eps$. 
\begin{prop}\label{lemB}
Under the condition $\lambda<\mu$, there exist some constants $K$ and $\eps_0>0$ such that
for any $\eps<\eps_0$ and $n\geq 1$, 
\[
\sup_{x\in{\cal S}} \E\left(\pB_n\mid X(0)=x\right)\leq Kn.
\]
\end{prop}
\begin{proof}
If one chooses $\eps_0$ so that 
\[
\mu_0\stackrel{\text{def.}}{=}\mu-\eps_0 \inf\{p^-(x): x\in{\cal S}\} >\lambda,
\]
then clearly the number of customers of the P-Queue is certainly smaller than the number
of customers of an $M/M/1$ queue with arrival rate $\lambda$ and service rate
$\mu_0$. Consequently, the corresponding busy periods compare in the same way, hence it is
enough to take $K=1/(\mu_0-\lambda)$. 
\end{proof}
The queue with time-varying service rate as defined above  will be
referred to as the perturbed queue, denoted, for short, by P-Queue. The case $\eps=0$
obviously corresponds to the S-Queue.   

\subsection{Adding and Canceling Departures}

The basic idea of the perturbation analysis carried out in this paper is to construct a
coupling of the busy periods of the processes $(\wL(t))$ and $(\pL(t))$. This is done as
follows, provided that for both queues the arrival process is ${\cal N}_\lambda$. 
\medskip

\paragraph{\bf Additional departures.} We denote by ${\cal N}^+$ the non-homogeneous Poisson process
whose intensity  is given by $t\to \eps p^+(X(t))$. Conditionally on $(X(t))$, the
number of points of ${\cal N}^+$ in the interval $[a,b]$, $0\leq a\leq b$ is Poisson with parameter
\[
\eps\int_a^b  p^+(X(s))\,ds.
\]
The points of ${\cal N}^+$ are denoted by $0<t^+_1\leq t_2^+\leq \ldots \leq t_n^+\leq \ldots$ 
and are called additional departures.
In particular the distribution of the location $t_1^+$ of the first point of ${\cal N}^+$
after $0$ is given by, for $x \ge 0$,
\begin{equation}\label{eqt1}
\P(t_1^+\geq x)= \P({\cal N}^+([0,x])=0)=\E\left(\exp\left(-\eps \int_0^x  p^+(X(s))\,ds\right)\right).
\end{equation}
See Grandell~\cite{Grandell:01} for an account on non-homogeneous Poisson processes,
referred to as doubly stochastic Poisson processes. 

\medskip
\paragraph{\bf Canceling Departures}

We denote by ${\cal N}^-$ the point process obtained by {\em thinning} the point process
${\cal N}_\mu$ (see Robert~\cite{Robert:08}). It 
is defined as follows: A point at $s>0$ of the
Poisson process ${\cal N}_\mu$  is a point of ${\cal N}^-$ with probability $\eps
p^-(X(s))  / \mu$. In this way, ${\cal N}^-$ is a stationary point process with intensity  
$\eps p^-(X(s))$. A point of ${\cal N}^-$ is called a canceled departure. The points of
the point process ${\cal N}^-$
are denoted by $0<t^-_1\leq t^-_2\leq \ldots \leq t^-_n \leq \ldots$. For $x>0$, by definition,
\begin{equation}\label{eqtb1}
\P(t^-_1\geq x)=\E\left(\prod_{i=1}^{{\cal N}_\mu([0,x])}\left(1-\frac{\eps p^-(X(s_i))}{\mu}\right)\right),
\end{equation}
where $(s_i)$ are the points of the point process ${\cal N}_\mu$.

With the above notation, it is not difficult to show that the Markov process $(\pL(t))$
has the same distribution as the solution of the stochastic differential equation  
\begin{equation}\label{Rdiff}
d\pL(t)=d\,{\cal N}_\lambda(t) -\ind{\pL(t-)>0} d \left({\cal N}_\mu+{\cal N}^+-{\cal N}^-\right)(t),
\end{equation}
which is the analogue of Equation~\eqref{Sdiff} for the P-Queue.

%Let $0<t^-_1\leq t^-_2\leq \cdots \leq t^-_n \leq \cdots$ denote the sequence of (potential) canceled departures
%which can be obtained  from the departures of the standard queue. If at time 0 both queues have the 
%same number of customers, then for $x>0$,
%\begin{equation}\label{eqtb1}
%\P(t^-_1\geq x)=\E\left(\prod_{i=1}^{{\cal D}([0,x])}\left(1-\frac{\eps p^-(X(D_i))}{\mu}\right)\right),
%\end{equation}
%where ${\cal D}=(D_i)$ is the point process of the departures from the S-Queue. 

%\begin{equation}\label{eqtb1}
%\P(t^-_1\geq x)=\E\left(\prod_{i=1}^{{\cal D}([0,x])}\left(1-\frac{\eps p^-(X(D_i))}{\mu}\right)\right),
%\end{equation}
%where ${\cal D}=(D_i)$ is the point process of the departures from the S-Queue. 

\section{Busy period analysis: First order term}\label{BPFsec}
Let us assume that  a busy period with one customer starts at time  $0$ in the S-Queue and
P-Queue. In  this section, we determine  the first term  of the power series  expansion in
$\eps$ of the expected value of $\pT$, the  duration of the busy period in the P-Queue. This
derivation allows  us in  addition to lay  down part  of the material  needed in  the next
section to compute the more intricate second term of the power series expansion in $\eps$.

For the first order term, we only have to consider the cases when there is either a single additional departure or else a single canceled departure. The probability that both events occur in the same busy period is clearly of the order of magnitude of $\eps^2$ since the intensities of the associated Poisson processes are proportional to $\eps$.

For $x\geq 1$, the stability assumptions ensure that the expected values of the busy periods starting
with $x$ customers, namely $\E(\wB_x)$ and $\E(\pB_x)$,  are both finite. 
When the first additional and canceled departures are such that $t^+_1>\pT$ and $t^-_1>\pT$ then
$\wT=\pT$. We now consider the different possibilities.
\subsubsection*{A single additional departure}

If there is only one additional departure and no canceled departure in $(0,\pT)$ then at time $\pT$, the
P-queue  is empty and the S-queue  is with one customer 
(see Figure~\ref{onedep}). 

\setlength{\unitlength}{1947sp}%
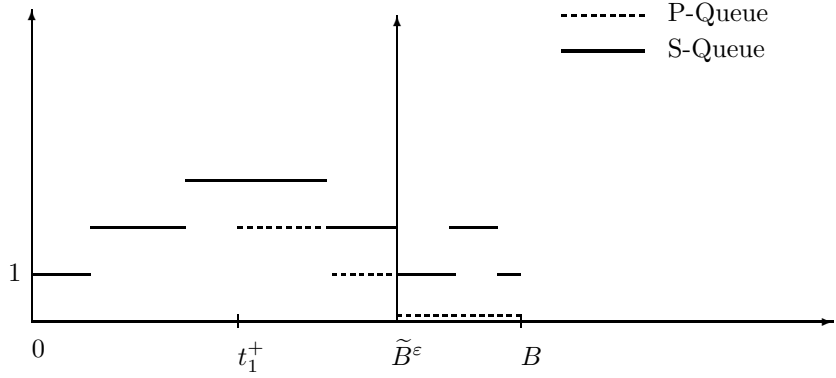
\begin{figure}[ht]
\begin{picture}(10512,4515)(1576,-5686)
\thinlines
{\put(1876,-5161){\vector( 0, 1){3975}}
\put(1876,-5161){\vector( 1, 0){10200}}
}%
\thicklines
{\put(1876,-4561){\line( 1, 0){750}}
}%
{\put(2626,-3961){\line( 1, 0){1200}}
}%
{\put(3826,-3361){\line( 1, 0){1800}}
}%
{\put(5626,-3961){\line( 1, 0){900}}
}%
{\put(6526,-4561){\line( 1, 0){750}}
}%
{\put(7201,-3961){\line( 1, 0){600}}
}%
{\put(7801,-4561){\line( 1, 0){300}}
}%
{\put(8101,-5161){\line( 1, 0){675}}
}%
\thinlines
{\multiput(4501,-3961)(123.52941,0.00000){9}{\line( 1, 0){ 61.765}}
}%
{\multiput(5701,-4561)(115.38462,0.00000){7}{\line( 1, 0){ 57.692}}
}%
{\multiput(6526,-5086)(116.66667,0.00000){14}{\line( 1, 0){ 58.333}}
}%
{\put(6526,-5161){\vector( 0, 1){3900}}
}%
{\multiput(8626,-1261)(123.52941,0.00000){9}{\line( 1, 0){ 61.765}}
}%
\thinlines
{\put(4501,-5086){\line( 0,-1){150}}
}%
{\put(8101,-5086){\line( 0,-1){150}}
}%
\thicklines
{\put(8626,-1711){\line( 1, 0){1050}}
}%
\thinlines

\put(8101,-5711){$\wT$}%
\put(8101,-5150){\line( 0, 1){75}}
\put(6451,-5711){$\pT$}%
\put(4501,-5711){$t_1^+$}%
\put(4501,-5150){\line( 0, 1){75}}
\put(1876,-5611){$0$}%
\put(1576,-4636){$1$}%
\put(9976,-1336){P-Queue}%
\put(9976,-1786){S-Queue}%
\end{picture}%
\caption{A busy Period with an Additional Departure}\label{onedep}
\end{figure}

We specifically prove the following lemma.

\begin{lem}
In the case of a single departure, we have
\begin{equation} \label{BP-1+}
\E\left((\wT-\pT)\ind{t^+_1 < \wT}\right)= \eps  \frac{\E_\nu[p (X(0))^{+}]}{(\mu-\lambda)^2}+o(\eps),
\end{equation} 
where $\nu$ is the equilibrium distribution of the environment $(X(t))$. 
\end{lem}

\begin{proof}
When there is only one additional departure, the variable $\pT$ is between $t^+_1$ and  $t^+_2$. We can write
%No points of ${\cal N}^-$ play a role in this case. Hence, the first point of the point process ${\cal N}^-$ satisfies $t^-_1>\pT$.  
\begin{equation}\label{aux0}
\E\left((\wT-\pT) \ind{t^+_1 < \wT}\right)= \E\left((\wT-\pT)\ind{t^+_1 < \pT <t^+_2,
  t^-_1> \pT}\right) + \Delta,
\end{equation}
where  the offset term $\Delta$ can be bounded as follows
\begin{equation}\label{aux1}
\Delta \leq \E\left(|\wT-\pT|\left( \ind{t^+_2 < \pT, t^-_1 > \pT} + \ind{t^-_1 \leq
    \pT, t^+_1 \leq    \pT }\right)\right).
\end{equation}

Let us estimate the first term of the right-hand side of ~\eqref{aux0}.
Equation~\eqref{eqt1} and the boundedness of $p$ give that
\begin{align*}
\P(t^+_1\leq \wT)&=1-\E\left(\exp\left(-\eps\int_0^{\wT} p^+(X(s))\,ds\right)\right)
\\&=\eps \E\left(\int_0^{\wT} p^+(X(s))\,ds\right)+o(\eps)\\
&=\eps \E(\wT)\,\E_\nu\left[p^+(X(0))\right]+o(\eps)=
\frac{\eps}{\mu-\lambda}\E_\nu\left[p^+(X(0))\right]+o(\eps),
\end{align*}
by independence between $\wT$ and $(X(t))$ and the stationarity of $(X(t))$. 
 By the strong Markov property at the stopping time $\pT$, conditionally on the event $\{t^+_1  < \pT <  t^+_2 , \pT < t^-_1 \}$, the S-Queue starts at $\pT$ an
  independent busy period with one customer, therefore
\begin{multline*}
\E\left((\wT-\pT)\ind{t^+_1 < \pT <t^+_2,  t^-_1> \pT}\right)=
\P(t^+_1 < \pT<t^+_2,  t^-_1> \pT)\\\times \E\left((\wT-\pT)\mid t^+_1 < \pT<t^+_2,  t^-_1> \pT\right)=
\P(t^+_1 < \pT<t^+_2,  t^-_1> \pT)\E(\wB_1).
\end{multline*}
% Now note that, by definition, 
% \begin{multline*}
% \{t^+_1 < \wT < t^+_2 , \wT < \bar t^+_1 \}\subset \{t^+_1 < T < t^+_2 , T < \bar t^+_1 \}\\
% \subset \{t^+_1 < \wT < t^+_2 , \wT < \bar t^+_1 \}\cup \{t^+_2 < \wT  \}
% \cup \{t^+_1 < \wT, \bar t^+_1< \wT \}.
% \end{multline*}
% With the same method as in the estimation of $\Delta$, one gets that
% \[
% \left|\E\left((\wT-T)\ind_{\{t^+_1 < T<t^+_2,  t^-_1\geq T\}}\right)
% -\E\left((\wT-T)\ind_{\{t^+_1 < \wT<t^+_2,  t^-_1\geq \wT\}}\right)\right|=o(\eps)
% \]
% and
% \[
% \left|\P\left(t^+_1 < T<t^+_2,  t^-_1\geq T\right)
% -\P\left(t^+_1 < \wT<t^+_2,  t^-_1\geq \wT\right)\right|=o(\eps). 
% \]
Now, since $\{t_1^+ < \pT\}=\{t_1^+ < \wT\}$ on the event $\{t^+_1  < \pT <  t^+_2 , \pT < t^-_1 \}$, then
\begin{align*}
\P(t^+_1 < \pT <t^+_2,  t^-_1> \pT)  &= \P(t^+_1 < \wT) - \P(t^+_1 < \pB,\, t^+_2 < \pT) - \\
&\P(t^+_1 < \pT,\, t^-_1 < \pT) + \P(t^+_1 < \pT,\, t^+_2 < \pT, \, t^+_1 < \pT) \\
&=\P(t^+_1 < \wT)  + o(\eps),
\end{align*}
since two or more extra jumps in the same busy period is $o(\eps)$.
Similarly, by using again the strong Markov property, one gets the following estimation
\begin{align*}
\E\left(|\wT-\pT|\ind{t^+_2 < \pT, t^-_1> \pT}\right)&\leq \sum_{n\geq 2}
\E(\wB_n)\P(t^+_n\leq \pT\leq t^+_{n+1},t^-_1\geq \pT)\\
&\leq  \frac{1}{(\mu-\lambda)}\sum_{n\geq2} n  \P(\cal{N}^+([ 0,\wT])=n).
\end{align*}
Indeed, given the S-Queue, $\cal{N}^+([ 0,\wT])$ has a Poisson distribution with
parameter $\int_0^{\wT} \eps  p^+(X(s))\,ds$, which implies that
\begin{multline*}
\sum_{n\geq2} n  \P(\cal{N}^+([ 0,\wT])=n) = 
\\ \E\left( \int_0^{\wT} \eps p^+(X(s))\,ds \right) - 
\E\left(\int_0^{\wT} \eps p^+(X(u))\,du  \, e^{-\eps \int_0^{\wT} p^+(X(s))\,ds}\right)= o(\eps)
\end{multline*}
and the first term in the right hand side of Inequality~\eqref{aux1} is thus negligible at the
first order in $\eps$. 
 
To estimate the second  term in the right hand side of Inequality~\eqref{aux1}, we need to
 consider the different possibilities for the location of the points $t_1^+$ and $t_1^-$. In the case that $t_1^+$ and $t_1^-$ occur during $[0,B]$ 
 and $\pT\geq  \wT$, at time $\wT$ the P-Queue  has at most $p\geq 0$ customers if there
 have been $p+1$ canceled departures. If ${\cal D}([0,\wT])$ is  the number of customers
 during the busy  period of the S-Queue, then certainly 
\begin{align*}
\E\left((\pT-\wT)\right.&\left.\ind{\pT\geq \wT, t^-_1 \leq    \wT, t^+_1 \leq    \wT  }\right)\\
&\leq \E\left(\E_{ X(\wT)}\left(B_{{\cal D}([0,\wT])}\right)\right) \P(t^-_1 <  \wT,
t^+_1\leq \wT \leq t^+_2)\\
&\leq K \E\left({\cal D}([0,\wT])\right)\P(t^-_1 <  \wT, t^+_1\leq \wT\leq t^+_2) =o(\eps),
\end{align*}
by Proposition~\ref{lemB}. On the other hand,
\begin{align*}
\E\left(\left|\pT-\wT\right|\right.&\left.\ind{\pT < \wT, t^-_1 \leq    \wT, t^+_1 \leq    \wT  }\right)
\leq \E\left(\wB\ind{t^-_1 \leq \wT, t^+_1\leq    \wT  }\right)
=o(\eps).
\end{align*}
Finally,
\begin{multline*}
\E\left(\left|\pT-\wT\right|\ind{t^-_1 \leq \pT, t^+_1 \leq    \pT  }\right)
\\\leq 
\E\left(\left|\pT-\wT\right|\ind{ t^-_1 \leq    \wT, t^+_1\leq    \wT  }\right)
+\E\left(\wB\ind{t^-_1 \leq \wT,  \wT  \leq t^+_1 \leq \pT   }\right),
\end{multline*}
where it can be shown in a similar way as before that the last term is  $o(\eps)$. One concludes that the term $\Delta$ is $o(\eps)$ as $\eps$ goes to $0$. By using Equation~\eqref{aux0}, we obtain the desired result.
\end{proof}
The estimation of the right hand side of Equation~\eqref{aux0} may appear quite
cumbersome. It is however worth noting that the environment $(X(t))$ of the P-Queue
introduces delicate dependences, which have to be handled  with care. This is why we have
chosen to explicitly write the precise setting in which  the strong Markov property is
used  to get the first order term.  In the following, similar arguments will not be explicitly formulated. 

\subsubsection*{A single canceled departure}
Suppose now that there is only one canceled departure, i.e. a departure of the S-Queue is canceled
for the P-Queue,  and no additional jumps during the busy period
of the S-Queue. In this case, at the end of the busy period of the S-Queue, at time $\wT$,
the P-queue has one customer and thus starts a busy period. Provided that there are no more canceled and
additional departures during $(\wT,\pT)$ in the P-Queue  then the difference between both busy periods
has the same distribution as the length $\wB_1$ of a standard busy period.  See
Figure~\ref{RD}. 

\bigskip 
\setlength{\unitlength}{1947sp}%
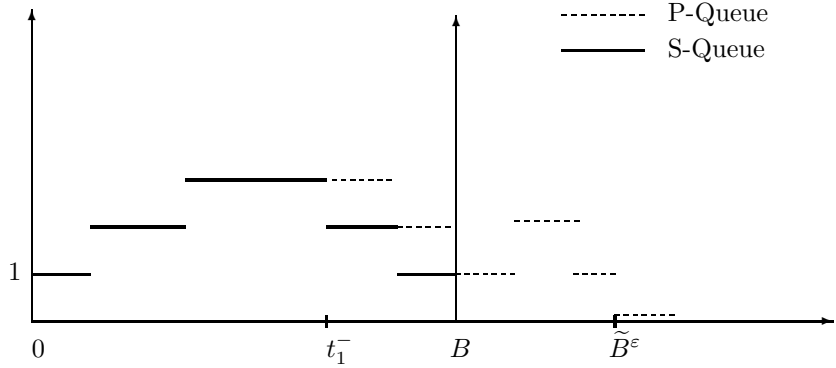
\begin{figure}[ht]
\begin{picture}(10512,4440)(1576,-5611)
\thinlines
{\put(1876,-5161){\vector( 0, 1){3975}}
\put(1876,-5161){\vector( 1, 0){10200}}
}%
\thicklines 
{\put(1876,-4561){\line( 1, 0){750}}}
{\put(2626,-3961){\line( 1, 0){1200}}
}%
{\put(3826,-3361){\line( 1, 0){1800}}
}%
{\put(5626,-3961){\line( 1, 0){900}}
}%
{\put(6526,-4561){\line( 1, 0){750}}
}%
\thinlines
{\multiput(8626,-1261)(123.52941,0.00000){9}{\line( 1, 0){ 61.765}}
}%
\thicklines
{\put(8626,-1711){\line( 1, 0){1050}}
}%
\thinlines
{\multiput(5701,-3361)(115.38462,0.00000){7}{\line( 1, 0){ 57.692}}
}%
{\multiput(7276,-4561)(115.38462,0.00000){7}{\line( 1, 0){ 57.692}}
}%
{\multiput(9301,-5086)(115.38462,0.00000){7}{\line( 1, 0){ 57.692}}
}%
\thicklines
{\put(7276,-5161){\line( 1, 0){675}}
}%
{\put(5626,-5086){\line( 0,-1){150}}
}%
{\put(9301,-5086){\line( 0,-1){150}}
}%
\thinlines
{\put(7276,-5161){\vector( 0, 1){3900}}
}%
{\multiput(6526,-3961)(122.72727,0.00000){6}{\line( 1, 0){ 61.364}}
}%
{\multiput(8026,-3886)(126.92308,0.00000){7}{\line( 1, 0){ 63.462}}
}%
{\multiput(8776,-4561)(116.66667,0.00000){5}{\line( 1, 0){ 58.333}}
}%
\put(1876,-5611){$0$}%
\put(1576,-4636){$1$}%
\put(9976,-1336){P-Queue}%
\put(5626,-5611){$t_1^-$}%
\put(7201,-5611){$\wT$}%
\put(9976,-1786){S-Queue}%
\put(9226,-5611){$\pT$}%
\end{picture}
\caption{A Busy Period with a Canceled Departure}\label{onemark}
\label{RD}
\end{figure}

\bigskip

\begin{lem}
In the case of a single canceled departure, we have
\begin{equation}\label{BP-1-}
\E\left((\pT   -\wT)\ind{t^-_1\leq \wT}\right) =
\eps  \frac{\E_\nu[p^-(X(0))]}{(\mu-\lambda)^2}+o(\eps).
\end{equation} 
\end{lem}

\begin{proof}
By using the same arguments as before, one obtains the relation
\begin{align*}
\E\left((\pT   -\wT)\ind{t^-_1\leq \wT}\right) &= 
\E\left(\wB_1 \ind{t^-_1\leq \wT,\,  \wT +\wB_{1} < \min(t^+_1, t^-_2)}\right) +o(\eps) \\
&= \E(\wB_1)\P(t^-_1\leq \wT)+o(\eps).
\end{align*}

To estimate $\P(t^-_1 \leq \wT )$, denote by $(D_i)$  the
sequence of departures times in the $S$-Queue and $N$ the number of customers served during the busy period
of length $\wT$, then Equation~\eqref{eqtb1} gives the identity
%To estimate $\P(t^-_1 \leq \wT )$, let $(D_i)$ denote the
%sequence of departures times in the $S$-Queue and $N$ the number of customers served during the busy period
%of length $\wT$, then Equation~\eqref{eqtb1} gives the identity
\begin{align*}
\P(t^-_1 \leq \wT ) &= \E \left(  \sum_{i=1}^{N} \frac{\eps
  p^{-}(X(D_{i}))}{\mu} \prod_{j=1}^{i-1} \left ( 1- \frac{\eps p^-(X(D_j))}{\mu} \right )
\right)\\
&=\frac{\eps}{\mu}  \E\left( \sum_{i=1}^{N}   p^-(X(D_{i})) \right) +o(\eps)\\
&=\frac{\eps}{\mu}  \E(N)\E\left (p^-(X(D_{1}))\right) +o(\eps)
\end{align*}
by stationarity of $(X(t))$ and Wald's Formula. Since $\E(N)=\mu/(\mu-\lambda)$ (see Appendix), Equation~\eqref{BP-1-} follows.
\end{proof}

In the expansion of the busy period of the P-Queue, the term in $\eps$ is given by the two
events consisting in only one canceled or only one additional departure during the busy
period of the S-queue. The next proposition follows from Equations~\eqref{BP-1+}
and~\eqref{BP-1-}. 
\begin{prop}[First Order Expansion]
\begin{equation}
\label{foxp}
\E (\pT) =  \frac{1}{\mu-\lambda} -\eps \frac{\E_\nu[p(X(0))]}{(\mu-\lambda)^2}  + o(\eps).
\end{equation} 
\end{prop} 

Equation~\eqref{foxp} is consistent with the so-called Reduced Service Rate
approximation. As a matter of fact, everything happens as if we had a classical $M/M/1$
queue with service rate $\mu+\eps\E_\nu[p(X(0))]$ and arrival rate $\lambda$. In that queue,
the mean length of the busy period is given by 
$$
\frac{1}{\mu+\eps\E_\nu[p(X(0))]-\lambda}= \frac{1}{\mu - \lambda}  -\eps \frac{\E_\nu[p(X(0))]}{(\mu-\lambda)^2}  + o(\eps),
$$
which coincides with Equation~\eqref{foxp}. In the following section, we investigate the second order term and show that the RSR approximation is no more valid.

\section{Busy Period: Second order term}\label{BPSterm}
In  this section,  the  coefficient  of $\eps^2$  of  the mean  busy  period $\E(\pT)$  is
calculated. In  the same way as  for the first order,  this coefficient is  related to the
event that  two extra jumps  occur during  a busy period  of the perturbed  $M/M/1$ queue.
Since extra  jumps can  be either additional  departures or canceled  departures, there
are three cases to investigate.  As it will be  seen, this coefficient stresses the
importance of the evolution  of the varying capacity, in
particular through its correlation function. This was not  the case for the  first order term,  since only the average
value of the capacity shows up there. 

In the following, in order to get the $\eps^2$ coefficient, one has to consider the
different possibilities for the location of the points $t^+_1$, $t^+_2$ and $t^-_1$,
$t^-_2$. By using similar arguments as in Section~\ref{BPFsec}, it is not difficult to
show that any event involving $t^+_3$ or $t^-_3$ yields a term of the order $\eps^3$ in
the expansion of $\E(\pT-\wT)$. 

Define
\[
{\cal A}_{+} = \{t^+_1\leq \wT, t^-_1\geq t^+_1+\wB_{\wL(t^+_1)-1}\}.
\]
On this event, at least one departure is added and the busy period of the P-Queue finishes before a
departure is canceled (note that $\wB_{\wL(t^+_1)-1}$ is the length of a busy period of S-Queue
starting at time $t_1$ with $L(t^+_1)-1$  customers).
On the event
\[
{\cal A}_{\pm} =\{t^-_1\leq \wT,\, \wT\leq t^+_1\leq  \wT +\wB_{1}\},
\]
a canceled departure occurs and another departure is added before the completion of the busy period of the P-queue, where $\wB_{1}$ denotes the duration of the additional busy period due to the canceled
departure. Finally, on the event 
\[
{\cal A}_{-}=\{t^-_1\leq \wT,\,  \wT +\wB_{1}\leq t^+_1\},
\]
at least a canceled departure occurs and no additional departures are added before the
completion of the busy period $\wT_1$.

By checking all the different cases, it is not difficult  to see that if ${\cal A}={\cal A}_{+}\cup {\cal A}_{\pm}\cup{\cal A}_{-}$,
the expression $\E((\pT-\wT)\mathbbm{1}_{{\cal A}^c})$ is $o(\eps^2)$ (and even equal to 0 in some cases, for instance when there are a canceled departure and an additional departure in such a way that $\pT=\wT$).  The following sections are
devoted to the estimation of $\E((\pT-\wT)\mathbbm{1}_{A})$ for $A\in \{{\cal A}_{+}, {\cal  A}_{\pm}, {\cal A}_{-}\}$.

In a first step, we analyze the case when there are only additional departures before $\wT$, that is, we consider the term $\E((\pT-\wT)\mathbbm{1}_{\mathcal{A}_+})$.
When no canceled departure occurs, at
most two additional departures in the time interval $[0,\wT]$, occurring at times $t^+_1$ and
$t^+_2$ respectively, may play a role in  the computation of  the coefficient of $\eps^2$ of $\E(\wT-\pT)$. 
In this case, the difference between $\wT-\pT$ is  equal to the busy period of an S-Queue which starts
with either one or two customers, depending on the fact that, on the event $\{t^+_1\leq \wT\}$,
the busy period of the P-Queue is already completed at time $t^+_2$ or not. See
Figure~\ref{twodep}.

As before, $\wB_2$  denotes a random variable with  the same  distribution  as  the sum  of  two
independent variables distributed as $\wB_1$ and independent of $\wT$, $t^+_1$ and $t^+_2$.
One gets
\begin{multline*}
\E \left((\wT-\pT)\mathbbm{1}_{{\cal A}_+} \right) = \E\left((\wT-\pT)\mathbbm{1}_{\{t^+_1\leq \wT,
  t^-_1\geq t^+_1+\wB_{\wL(t^+_1)-1}\}}\right)\\
=  \E\left(\wB_2\right)\P\left(t^+_1 < \wT, t^+_2 < t^+_1 + \wB_{\wL(t^+_1) -1}, t^-_1\geq t^+_1+\wB_{\wL(t^+_1)-1}\right) \notag
\\+ \E\left(\wB_1\right)\P\left(t^+_1 < \wT, t^+_2\geq t^+_1 + \wB_{\wL(t^+_1) -1},
  t^-_1\geq t^+_1+\wB_{\wL(t^+_1)-1}\right) + o(\eps^2).
\end{multline*}
This decomposition entails that
\begin{multline}\label{bp21}
\E\left((\wT-\pT)\mathbbm{1}_{{\cal A}_+} \right) = \left(\E\left(\wB_2\right)-\E\left(\wB_1\right)\right)\P\left(t^+_1 < \wT, t^+_2 < t^+_1 + \wB_{\wL(t^+_1) -1}\right) 
\\    +\E\left(\wB_1\right)\left(\P\left(t^+_1 < \wT\right)-\P\left(t^+_1 < \wT, t^-_1\leq t^+_1+\wB_{\wL(t^+_1)-1}\right)\right)+
o(\eps^2).
\end{multline}

\setlength{\unitlength}{1947sp}%
\begingroup\makeatletter\ifx\SetFigFont\undefined%
\gdef\SetFigFont#1#2#3#4#5{%
  \reset@font\fontsize{#1}{#2pt}%
  \fontfamily{#3}\fontseries{#4}\fontshape{#5}%
  \selectfont}%
\fi\endgroup%
\begin{figure}[ht]
\begin{picture}(10512,4440)(1576,-5611)
\thinlines
{\put(1876,-5161){\vector( 0, 1){3975}}
\put(1876,-5161){\vector( 1, 0){10200}}
}%
\thicklines
{ \put(1876,-4561){\line( 1, 0){750}}
}%
\thinlines
{\multiput(8626,-1261)(123.52941,0.00000){9}{\line( 1, 0){ 61.765}}
}%
\thicklines
{\put(8626,-1711){\line( 1, 0){1050}}
}%
\thinlines
{\put(6601,-5161){\vector( 0, 1){3900}}
}%
{\multiput(6601,-5086)(123.52941,0.00000){9}{\line( 1, 0){ 61.765}}
}%
{\multiput(5626,-4561)(114.70588,0.00000){9}{\line( 1, 0){ 57.353}}
}%
\thicklines
{\put(2626,-3961){\line( 1, 0){675}}
}%
\thinlines
{\multiput(4876,-3961)(126.92308,0.00000){7}{\line( 1, 0){ 63.462}}
}%
{\multiput(4276,-3361)(109.09091,0.00000){6}{\line( 1, 0){ 54.545}}
}%
\thinlines
{\put(4276,-5086){\line( 0,-1){150}}
}%
{\put(4876,-5086){\line( 0,-1){150}}
}%
{\put(10726,-5086){\line( 0,-1){150}}
}%
\thicklines
{\put(3301,-3361){\line( 1, 0){600}}
}%
{\put(5701,-3361){\line( 1, 0){900}}
}%
{\put(7351,-3361){\line( 1, 0){600}}
}%
{\put(6601,-3961){\line( 1, 0){750}}
}%
{\put(7951,-3961){\line( 1, 0){300}}
}%
{\put(9001,-3961){\line( 1, 0){750}}
}%
{\put(8326,-4561){\line( 1, 0){675}}
}%
{\put(9826,-4561){\line( 1, 0){825}}
}%
{\put(10726,-5161){\line( 1, 0){675}}
}%
{\put(3901,-2761){\line( 1, 0){1800}}
}%
\thinlines

\put(1876,-5611){0}%
\put(1576,-4636){1}%
\put(4151,-5736){$t_1^+$}%
\put(10500,-5736){$\wT$}%
\put(9976,-1711){S-Queue}%
\put(4751,-5736){$t_2^+$}%
\put(6501,-5736){$\pT$}%
\put(9976,-1336){P-Queue}%
\end{picture}
\caption{Two Additional Departures}\label{twodep}
\end{figure}
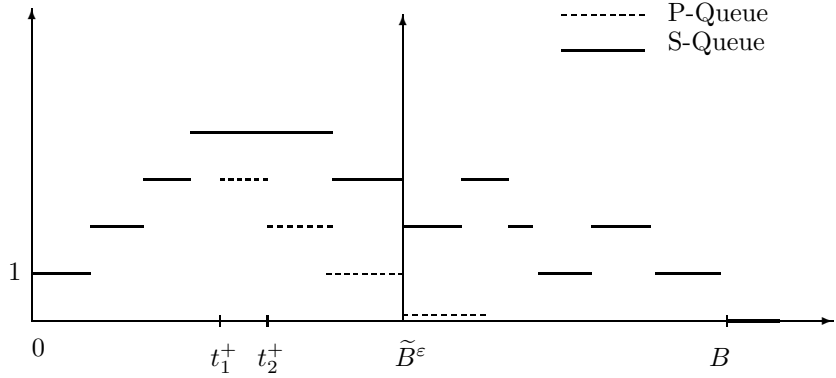

From Equation~\eqref{bp21}, one has to expand three expressions with respect to $\eps$. This is done by proving the three following lemmas.

\begin{lem}\label{expansion1}
The quantity $\P\left(t^+_1 < \wT, t^+_2 <t^+_1 + \wB_{\wL(t^+_1) -1}\right)$ can be
expanded as 
\begin{multline}\label{OK1}
\P\left(t^+_1 < \wT, t^+_2 <t^+_1 + \wB_{\wL(t^+_1) -1}\right)\\
=\rho\eps^2E\left(\int_0^{\wT}(\wT-v)\E_\nu\left(p^+(X(0))p^+(X(v))\right)\right)\,dv+o(\eps^2).
\end{multline}
\end{lem}

\begin{proof}
Let us recall the regenerative description of  a busy period starting at time $0$ with one
customer:  At time $E_1$  (exponentially distributed  with parameter  $\lambda+\mu$), with
probability $\mu/(\lambda+\mu)$  the busy period is finished.  Otherwise, with probability
$\lambda/(\lambda+\mu)$,  a  new  customer  arrives  and a  sub-busy  period  of  duration
$\wB_1^1$ (with  the same distribution  as $\wB_1$) begins  until the number  of customers
reaches $1$ again. In this way, the variable $\wT$ can be represented as follows
\begin{equation}
\label{represent}
\wT=E_0+\sum_{i=1}^H \left(E_i+\wB_1^i\right),
\end{equation}
where $H$ is geometrically distributed with parameter $\lambda/(\lambda+\mu)$, $(E_i)$ are
i.i.d exponentially distributed with parameter $\lambda+\mu$ and $(\wB_1^i)$ are
i.i.d. All these random variables are independent. For $0\leq i\leq H$, 
\begin{itemize}
\item[---] $s_i$ denotes the end of the $i$th sub-busy cycle:
$s_0=0$ and, for $i\geq 1$, $s_i=s_{i-1}+ E_{i}+\wB_1^i$, $B=s_{H}+E_0$;
\item[---]  $N_i$ denotes the number of arrivals   during the $i$th sub-busy cycle;
\item[---]  $s_{i-1}+D_1^i$, \ldots, $s_{i-1}+D_{N_i}^i$ are the instants of departures of customers
  during the $i$th sub-busy cycle.  
\end{itemize}
For the joint distribution of the vector $(N_i, D_1^i,\ldots, D_{N_i}^i)$, see the Appendix. Figure~\ref{bp4fig} gives an illustration of the above definitions.

\setlength{\unitlength}{2047sp}%
\begingroup\makeatletter\ifx\SetFigFont\undefined%
\gdef\SetFigFont#1#2#3#4#5{%
  \reset@font\fontsize{#1}{#2pt}%
  \fontfamily{#3}\fontseries{#4}\fontshape{#5}%
  \selectfont}%
\fi\endgroup%
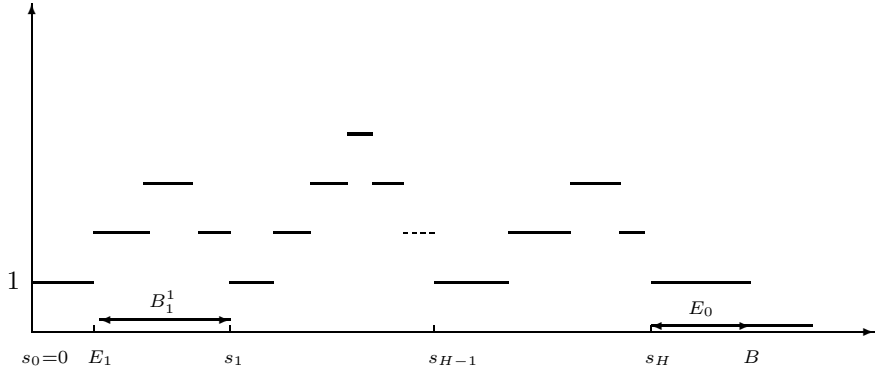
\begin{figure}[ht]
\begin{picture}(10512,4362)(1576,-5536)
\thinlines
{  \put(2626,-5161){\line( 0, 1){ 75}}
}%
{  \put(2701,-5011){\vector(-1, 0){  0}}
\put(2701,-5011){\vector( 1, 0){1575}}
}%
\thicklines
{  \put(2626,-3961){\line( 1, 0){675}}
}%
{  \put(3901,-3961){\line( 1, 0){375}}
}%
{  \put(4276,-4561){\line( 1, 0){525}}
}%
{  \put(4801,-3961){\line( 1, 0){450}}
}%
{  \put(5251,-3361){\line( 1, 0){450}}
}%
{  \put(3226,-3361){\line( 1, 0){600}}
}%
{  \put(5701,-2761){\line( 1, 0){300}}
}%
{  \put(6001,-3361){\line( 1, 0){375}}
}%
\thinlines
{  \multiput(6376,-3961)(107.14286,0.00000){4}{\line( 1, 0){ 53.571}}
}%
\thicklines
{  \put(6751,-4561){\line( 1, 0){900}}
}%
{  \put(7651,-3961){\line( 1, 0){750}}
}%
{  \put(8401,-3361){\line( 1, 0){600}}
}%
{  \put(9001,-3961){\line( 1, 0){300}}
}%
\thinlines
{  \put(1876,-5161){\vector( 0, 1){3975}}
\put(1876,-5161){\vector( 1, 0){10200}}
}%
{  \put(6751,-5161){\line( 0, 1){ 75}}
}%
{  \put(9376,-5086){\line( 0,-1){ 75}}
}%
\thicklines
{  \put(1876,-4561){\line( 1, 0){750}}
}%
{  \put(9376,-4561){\line( 1, 0){1200}}
}%
{  \put(10576,-5086){\line( 1, 0){750}}
}%
\thinlines
{  \put(9376,-5086){\vector(-1, 0){  0}}
\put(9376,-5086){\vector( 1, 0){1200}}
}%
{  \put(4276,-5161){\line( 0, 1){ 75}}
}%
\put(1576,-4636) {$1$}%
\put(3301,-4861) {$\scriptstyle B_1^1$}%
\put(1751,-5536) {$\scriptstyle s_0=0$}%
\put(2551,-5536) {$\scriptstyle E_1$}%
\put(6676,-5536) {$\scriptstyle s_{H-1}$}%
\put(9301,-5536) {$\scriptstyle s_H$}%
\put(10490,-5536) {$\scriptstyle \wT$}%
\put(9826,-4936) {$\scriptstyle E_0$}%
\put(4201,-5536) {$\scriptstyle s_1$}%
\end{picture}%
\caption{Decomposition of a Busy Period}\label{bp4fig}
\end{figure}

It is easy to see that for the event $\{t^+_1 \leq \wT\, , t^+_2 <t^+_1 + \wB_{\wL(t^+_1) -1}\}$ to occur, $t^+_1$ and $t^+_2$ have to be in the same sub-busy period,  $[s_{i-1}+E_{i},s_i]$, for some $i\in\{1,\ldots ,H\}$. For a fixed $i$, the probability that the first two additional jumps  are in the $i$th  sub-busy period, is 
\begin{align*}
& \E \left( \int_{s_{i-1}+E_i}^{s_i}  \eps p^+(X(u))
e^{-\eps\int_0^{u} p^+(X(s)) \,ds} \left (1-e^{-\eps\int_{u}^{s_i} p^+(X(s)) \,ds}\right )\, du \right )\\
&= \eps^2 \E \left( \int_{s_{i-1}+E_i}^{s_i} p^+(X(u))\int_{u}^{s_i} p^+(X(s)) \,ds \, du \right)
+o(\eps^2).
\end{align*}

Since,  $\wB_1^i=s_i-s_{i-1}- E_{i-1}$ has the same distribution as $\wT$ and  by  the stationarity of $((X(t))$, the coefficient of $\eps^2$ can be expressed as follows, 
\begin{multline*}
\E\left(\int_{0\leq u\leq  v\leq \wT} p^+(X(u))p^+(X(v))\,du\,dv\right)\\
=\E\left(\int_{0\leq u\leq v\leq \wT} \E_\nu\left[p^+(X(0))p^+(X(v-u))\right]\,du\,dv\right). 
\end{multline*}

Finally, since $H$ is geometrically distributed with parameter $\lambda/(\lambda+\mu)$, Equation 
\eqref{OK1} follows.
\end{proof}

We turn now to the expansion of the quantity $\P(t^+_1\leq \wT)$, which is of course a refinement of what has been done in Section~\ref{BPFsec}.

\begin{lem}
The quantity $\P(t^+_1\leq \wT)$ can be expanded as
\begin{multline}\label{OK2}
\P(t^+_1\leq \wT)=
\eps \frac{\E_\nu\left[p^+(X(0))\right]}{\mu-\lambda}\\-\eps^2\E\left(\int_{0}^{\wT}(\wT-v)\, \E_\nu\left(p^+(X(0))p^+(X(v))\right)\,dv\right)+o(\eps^2).
\end{multline}
\end{lem}

\begin{proof}
We clearly have
\begin{align*}
\P(t^+_1\leq \wT)&=\E\left(1-e^{-\eps\int_0^{\wT} p^+(X(s))\,ds}\right)\\
&= \eps \frac{\E_\nu\left[p^+(X(0))\right]}{\mu-\lambda}-\frac{\eps^2}{2}\E\left(\left(\int_0^{\wT} p^+(X(s))\,ds\right)^2\right)+o(\eps^2).
\end{align*}
The second moment of the integral can be expressed as follows, by symmetry,
\begin{align*}
\E\left(\left(\int_0^{\wT} p^+(X(s))\,ds\right)^2\right)&=2\E\left(\int_{0\leq u\leq
  v\leq \wT} p^+(X(u))p^+(X(v))\,du\,dv\right)\\
&=2\E\left(\int_{0\leq u\leq
  v\leq \wT} \E_\nu\left(p^+(X(0))p^+(X(v-u))\right)\,du\,dv\right),
\end{align*}
by stationarity of the process $(X(t))$ and Equation \eqref{OK2} follows.
\end{proof}

Finally, we examine the expansion of $\P(t^+_1 < \wT, t^-_1\leq t^+_1+\wB_{\wL(t^+_1)-1})$.
This term is more delicate to expand, because of the canceled departure.
\begin{lem}
The quantity $\P(t^+_1 < \wT, t^-_1\leq t^+_1+\wB_{\wL(t^+_1)-1})$ can be expanded as
\begin{multline}\label{OK3}
\P(t^+_1 < \wT, t^-_1\leq t^+_1+\wB_{\wL(t^+_1)-1})\\= 
\frac{\eps^2}{\mu} \E\left(\sum_{i=1}^{H} \sum_{j=1}^{N_i}\int_{0}^{A_i} p^+(X(u)) p^-(X(D_j^i))\,du \right)+o(\eps^2),
\end{multline}
where $H$ is geometric distributed with parameter $\lambda/(\mu+\lambda)$, $(N_i,D_1^i,\ldots D_N^i)$
denotes the number of departures and the departures times in a busy period of length $B^i$, and
$$A_i=B_1^i+E_0+\sum_{k=i+1}^{H} (E_k+B_1^k),$$
where $(E_i)$ are i.i.d is exponentially distributed with parameter $\mu+\lambda$ and $(B_1^i)$ are i.i.d with the same distribution as $\wT$.
\end{lem}
 
\begin{proof}
Using the regenerative description of a standard busy period introduced in the proof of Lemma \ref{expansion1}, the variable $t^-_1$ has to occur in some sub-busy period $[s_{i-1}+E_i,s_{i}]$ of $B$ for some $1 \leq i \leq H$. A little thought show that if $t^-_1\in [s_{i-1}+E_i,s_{i}]$ then $t^+_1$ has to be  in $[s_{i-1}+E_i,B]$ for the event $\{t_1^+ < B, t^-_1 \leq t^+_1+\wB_{\wL(t^+_1)-1}\}$ to occur. 
The probability that $t^-_1$ and $t_1^+$ are located in $[s_{i-1}+E_i,s_{i}]$
and $[s_{i-1}+E_i,B]$, respectively, is
\begin{multline*}
\E\left(\int_{s_{i-1}+E_i}^B \eps p^+(X(u)) e^{-\eps\int_0^{u} p^+(X(s)) \,ds }  \, du
\sum_{j=1}^{N_i}\eps\frac{p^-(X(s_{i-1}+D_j^i))}{\mu} 
\right .\\
\left .
\prod_{k=1}^{j-1} \left(1-\eps\frac{p^-(X(s_{i-1}+D_k^i))}{\mu}\right)
\prod_{l=1}^{i-1}\prod_{r=1}^{N_l} \left(1-\eps\frac{p^-(X(s_{l-1}+D_r^l))}{\mu}\right)
\right),
\end{multline*}
where the coefficient of $\eps^2$ is
$$
\frac{1}{\mu} \E\left(\sum_{j=1}^{N_i}\int_{s_{i-1}+E_i}^{\wT} p^+(X(u)) p^-(X(s_{i-1}+D_j^i))\,du \right).
$$
Considering the different sub-cycles during $B$ and by the stationarity of $(X(t))$, Equation~\eqref{OK3} follows.
\end{proof}

We are now able to compute the coefficient of $\eps^2$ in the power series expansion of
$\E((\pT-\wT)\mathbbm{1}_{\mathcal{A}_+})$ in $\eps$. 

\begin{prop}
The coefficient of $\eps^2$ in the  expansion of $\E((\wT - \pT)\mathbbm{1}_{\mathcal{A}_+})$ with
respect to  $\eps>0$ is given by
\begin{multline}
\label{a+}
a_+ = - \frac{1}{\mu}\E\left(\int_{0}^{\wT}(\wT-v)\, \E_\nu\left(p^+(X(0))p^+(X(v))\right)\,dv\right) \\
-\frac{1}{\mu^2(1-\rho)}  \E\left(\sum_{i=1}^{H} \sum_{j=1}^{N_i}\int_{0}^{A_i} p^+(X(u)) p^-(X(D_j))\,du \right). 
\end{multline}
\end{prop}
To complete the analysis, we now turn to the expansion of
$\E((\pT-\wT)\mathbbm{1}_{\mathcal{A}_\pm})$ and $\E((\pT-\wT)\mathbbm{1}_{\mathcal{A}_-})$. In the
calculations, it appears more convenient to consider the sum of both terms and we then
have the following result. 
\begin{prop}
The coefficient of $\eps^2$ in the expansion of
$\E((\pT-\wT)\mathbbm{1}_{\mathcal{A}_\pm\cup\mathcal{A}_-} )$ with respect to $\eps>0$ is given by
\begin{multline}
\label{a-}
a_- = \frac {1 }{\mu^2(1-\rho)} \left ( -  \E \left ( \sum_{i=1}^N   \int_0^{\wT + \wB_1} p^- (X(D_i))  p^+ (X(s))  \, ds \right  ) \right. \\ 
\left.+\frac{1}{\mu} \E\left (\sum_{i=1}^{N} \sum_{k=1}^{N^{\prime}}   p^- (X(0)) p^- (X(\wT -D_i + D_k^{\prime} ))\right ) \right ),
\end{multline}
where $(N,D_{1},\ldots,D_{N})$ and $(N',D_{1}',\ldots,D_{N'}')$ denote the number of
departures and the departure   times in the busy periods of length $\wT$ and $\wB_{1}$,
respectively.  
\end{prop}

\begin{proof}
When a single canceled departure occurs (at time $t^-_1$)  before  $\wT$, an additional busy
period of length $\wB_1$ has to be added to take into account the canceled departure. 

By the strong Markov property, with the same method as in Section~\ref{BPFsec}, one obtains the relation
\begin{multline*} 
\E\left((\wT + \wB_1  -\pT) \ind{t^-_1\leq \wT,\, \wT\leq t^+_1\leq  \wT +\wB_{1}}\right) 
 \\= \E\left(\wB_1'\right) P \left(t^-_1\leq \wT,\, \wT\leq t^+_1\leq  \wT +\wB_{1}
 \right) +o(\eps^2), 
\end{multline*} 
where the random variable $\wB_1'$ has the same distribution as the random variable $\wB_1$,
hence,
\begin{multline}
\label{bp22}
\E\left((\pT-\wT)\mathbbm{1}_{{\cal A}_{\pm}} \right) = \E\left((\pT-\wT)\ind{t^-_1\leq \wT,\, \wT\leq t^+_1\leq  \wT +\wB_{1}}\right) \\
=  \E\left(\wB_1  \ind{t^-_1\leq \wT,\, \wT\leq t^+_1\leq  \wT +\wB_{1}}\right) -\E\left(\wB_1'\right) \P \left(t^-_1\leq \wT,\, \wT\leq t^+_1\leq  \wT +
\wB_{1} \right) +o(\eps^2).
\end{multline}

Now, two canceled departures in the same busy period gives two additional independent busy
periods starting with one customer,
\begin{multline*}
\E\left((\pT-\wT)\mathbbm{1}_{{\cal A}_-} \right) =\E\left((\pT-\wT)\ind{t^-_1\leq \wT,\,  \wT +\wB_{1}\leq t^+_1}\right)\\=
\E\left(\wB_1 \ind{t^-_1\leq \wT,\,  \wT +\wB_{1}\leq \min(t^+_1, t^-_2)}\right) \\ +
\E\left(\left(\wB_1+\wB_1'\right) \ind{t^-_1\leq \wT,\,
\wT\leq  t^-_2\leq \wT +\wB_{1},\,  \wT +\wB_{1}+\wB_{1}'\leq t^+_1}\right)
\\+\E\left(\wB_2 \ind{t^-_1\leq \wT,\,  t^-_2\leq \wT,\,  \wT +\wB_{1}+\wB_{1}'\leq t^+_1}\right)+ o(\eps^2).
\end{multline*}
Hence,
\begin{multline*}
\E\left((\pT-\wT)\mathbbm{1}_{{\cal A}-} \right) =
\E\left(\wB_1  \ind{t^-_1 \leq  \wT, t^-_2 >  \wT +\wB_{1}}  \right)
-\E\left(\wB_1 \ind{t^-_1\leq \wT, t^+_1\leq \wT +\wB_{1}}\right) \nonumber \\
 +\E\left(\wB_1 \ind{t^-_1\leq \wT,\,\wT\leq  t^-_2\leq \wT +\wB_{1}}\right)
+\E\left(\wB_1'\right) \E\left(\ind{t^-_1\leq \wT,\,\wT\leq  t^-_2\leq \wT +\wB_{1}}\right)
\\+\E\left(\wB_2 \ind{t^-_2\leq \wT }\right)+ o(\eps^2).
\end{multline*}
Finally,
\begin{multline}\label{bp23}
\E\left((\pT-\wT)\mathbbm{1}_{{\cal A}_-} \right) = \E\left(\wB_1\right)\P\left(t^-_1 \leq  \wT, t^-_2 >  \wT  \right) -\E\left(\wB_1 \ind{t^-_1\leq \wT, t^+_1\leq \wT +\wB_{1}}\right) 
\\  + \E\left(\wB_1'\right) \P\left( t^-_1\leq \wT,\,\wT\leq  t^-_2\leq \wT
  +\wB_{1}\right)   +2\E\left(\wB_1\right)\P\left(t^-_2\leq \wT\right)  + o(\eps^2), 
\end{multline}

From Section~\ref{model}, it is not difficult to see that the expression
\[
\P\left(t^-_1 \leq  \wT, t^-_2 >  \wT\right)+ 2\P\left(t^-_2\leq \wT\right)
\]
has no term in $\eps^2$ in its power series expansion. Thus the first term and the last
term of the right hand side of Equation~\eqref{bp23} cancel out for the expansion. 

The following expansions are obtained in a similar way,
\begin{multline*}
\E\left( \wB_1  \ind{t^-_1\leq \wT,\, t^+_1\leq  \wT +\wB_{1}} \right) \\= \frac{\eps^2}{\mu}
 \E \left  ( \wB_1 \sum_{i=1}^N   
\int_0^{\wT + \wB_1} p^- (X(D_i))  p^+ (X(s))  \, ds \right  ) + o(\eps^2),
\end{multline*} 
and
\begin{multline*}
\P\left(t^-_1\leq \wT, \wT < t^-_2\leq \wT + \wB_{1}\right) \\= \frac{\eps^2}{\mu^2 } \E         \left (\sum_{i=1}^{N} \sum_{k=1}^{N^{\prime}}   p^- (X(0)) p^- (X(\wT -D_i + D_k^{\prime} )) \right )  + o(\eps^2),
\end{multline*}
where $(N,D_{1},\ldots,D_{N})$ and $(N',D_{1}',\ldots,D_{N'}')$ denote the number of
departures and the departures   times in two independent busy periods of lengths $\wT$ and $\wB_{1}$,
respectively.

If we sum up the expansions obtained for canceled departures and one canceled and one
additional departures  (Equations~\eqref{bp22} and~\eqref{bp23}), with standard
manipulations, one gets the second term of the expansion $\E((\pT-\wT)( \mathbbm{1}_{{\cal
    A}_{\pm}} + \mathbbm{1}_{{\cal A}_-} ))$ in $\eps$. 
\end{proof}

To summarize the results obtained in this section, we can state the following theorem.

\begin{theo}
The coefficient of $\eps^2$ is the power series expansion of $\E(\pT-\wT)$ in $\eps$  is
equal to $a_--a_+$, where the coefficients $a_+$ and $a_-$ are given by
Equations~\eqref{a+} and \eqref{a-}, respectively. 
\end{theo}
It should be noted that the distributions involved in Equations~\eqref{a+} and \eqref{a-}
can be explicited by using the classical results concerning the $M/M/1$ queue. See the
Appendix where they are recalled. In the next section, we examine some applications of
the above result. 

\section{Applications}\label{applications}
\subsection{Non-Negative Perturbation Functions} 
Equations~\eqref{OK1} and~\eqref{OK2} give that the expansion
\[
\E\left(\wT-\pT \right)= \delta_1\eps+\delta_2\eps^2 +o(\eps^2)
\]
holds, with
$\delta_1= {\E_\nu\left(p(X(0))\right)}/{(\mu-\lambda)^2}$ 
and
\[
\delta_2= -\frac{1}{\mu} \E\left(\int_{0}^{\wT} (\wT-v)\, \E_\nu\left(p(X(0))p(X(v))\right)\,dv\right).
\]

Denote by $C_p(u)=\E_\nu\left[p(X(0))p(X(u))\right]-\E_\nu\left[p(X(0))\right]^2$, 
the covariance of the extra capacity. The second term of the expansion can be expressed  as  
\[
\delta_2= -\frac{1}{\mu} \E\left(\int_{0}^{\wT}(\wT-v)C_p(v)\,dv\right)
-\frac{\E_\nu\left[p(X(0))\right]^2}{(\mu-\lambda)^3},
\]
hence,
\begin{multline*}
\E\left(\wT-\pT \right)=\eps\, \frac{\E_\nu\left[p(X(0))\right]}{(\mu-\lambda)^2}
-\eps^2\, \frac{\E_\nu\left[p(X(0))\right]^2}{(\mu-\lambda)^3}\\
-\frac{\eps^2}{\mu}\E\left(\int_{0}^{\wT}  \,(\wT-v) C_p(v)\,dv\right)+o(\eps^2).
\end{multline*}
The following proposition which readily follows, compares the length of the busy period of
the P-Queue with an $M/M/1$ queue with  service rate $\mu+\eps \E_\nu(p(X(0)))$.
\begin{prop}[Comparison with reduced service rate]\label{chevalier}
If $\hT$ is the length of a busy period of an $M/M/1$ queue with service rate
$\mu+\eps \E_\nu(p(X(0)))$ then
\[
\lim_{\eps\to 0}\frac{1}{\eps^2}\E\left(\hT-\pT \right)=-\frac{1}{\mu}\E\left(\int_{0}^{\wT} (\wT-v)\,C_p(v)\,dv\right),
\]
where,
for $u\geq 0$,
$$C_p(u)=\E_\nu\left[p(X(0))p(X(u))\right]-\E_\nu\left[p(X(0))\right]^2$$ is, up to the
factor $\eps^2$, the covariance function of the extra-capacity of the perturbed queue.  
\end{prop}
It is straightforward to conclude  from the expression in Proposition~\ref{chevalier} that
$\E(\hT-\pT)$ is  negative when  $\eps$ is small. 
\begin{corol}[Negative impact of the variation of the service rate]  
When the environment is positively correlated i.e. when the function $u \rightarrow
C_p(u)$ is non-negative, then the first term of the expansion of $\E(\hT-\pT)$ in $\eps$ is of order $2$ and is negative.
\end{corol}
The following expression gives a closed form expression of the second term of the
expansion when the environment has an exponential decay.

\begin{prop}
When the correlation function of the
environment is exponentially decreasing, i.e. when, for some $\alpha >0$,
$$C_p(x)=\Var[p(X(0))]\,e^{-\alpha x},
\qquad x\geq 0,$$  then the difference between reduced  and variable
service rates satisfies the relation
\begin{equation}
\lim_{\eps\to 0} \frac{1}{\eps^2}\,\E\left(\hT-\pT \right)
\stackrel{\text{def.}}{=}\Delta_2(\alpha)=
-\,\frac{\Var[p(X(0))]}{(\mu-\lambda)^3}\, \E\left(e^{-\alpha Z}\right)\leq 0,
\end{equation}
where, $Z$ is the random variable whose density function on $\R_+$ is given by
\[
x \to \frac{1}{\mu(1-\rho)^2}  \int_{x}^{+\infty} \P\left(\wT\geq u\right)\,du.
\]
In particular, the function $\alpha\to \Delta_2(\alpha)$ is non-decreasing and concave. 
\end{prop}
\begin{proof}
For a square integrable random variable $A$ on $\R^+$, $A^*$ denotes the random variable
with density  $x \to \P(A\geq u)/{\E(A)}$
on $\R_+$. Note that for $\alpha\geq 0$, 
\begin{equation}\label{recaux}
\E\left(e^{-\alpha A^*}\right)=\frac{1-\E(e^{-\alpha A})}{\alpha \E(A)}
\end{equation}
and $\E(A^*)=\E(A^2)/(2\E(A))$.

To simplify notations, it is assumed that $\Var[p(X(0))]=1$. 
Proposition~\ref{chevalier} gives that the coefficient $\Delta_2(\alpha)$ of $\eps^2$ is in this case 
\begin{align*}
\Delta_2(\alpha)
&=-\frac{1}{\mu}\E\left(\int_{0}^{\wT}\left(\wT-v\right)\,e^{-\alpha v}\,dv\right)=-\frac{1}{\mu}\E\left(\int_{0}^{\wT}v\,e^{-\alpha (\wT-v)}\,dv\right)\\
&=-\frac{1}{\mu}\E\left(\frac{\wT}{\alpha}-\frac{1}{\alpha^2}+\frac{e^{-\alpha
    \wT}}{\alpha^2}\right)
=-\frac{\E(\wT)\E(\wT^*)}{\mu}\frac{1-\E\left(e^{-\alpha\wT^*}\right)}{\alpha\E(\wT^*)}.
\end{align*}
The Proposition is proved by using Relation~\eqref{recaux}. 
\end{proof}
\subsection{Non-Positive Perturbation Functions} 
It is assumed in  this section that the perturbation function is  non-positive so that the
environment  uses a part of the capacity  of  the $M/M/1$  queue with  constant service  rate
$\mu$. This application is motivated by  the following practical situation: Coming back to
the coexistence of elastic and streaming  traffic in the Internet, assume that priority is
given  to streaming  traffic in  a buffer  of a  router. The  bandwidth available  for
non-priority  traffic  is  the  transmission  link  reduced  by  the  bit  rate  of
streaming 
traffic. Denoting  by $\eps d(X_t)$  the bit  rate of streaming  traffic at time  $t$ (for
instance $\eps$ may represent the peak rate of a streaming flow and $d(X_t)$ the number of
such flows active  at time $t$), the  service rate available for non-priority traffic is
$\mu -  \eps d(x)$.    Setting $p(x) = - d(x)$, the
function $p(x)$ is non-positive.  We are  then in the framework when the environment gives
a reduced bandwidth  to a non-priority $M/M/1$  queue.
The same notation as in the previous section is  used extensively.

Equations~\eqref{BP-1+} and~\eqref{a-} give that the expansion
\[
\E\left(\wT-\pT \right)= \delta_1\eps+\delta_2\eps^2 +o(\eps^2)
\]
holds, with
$\delta_1= {\E\left[p(X(0))\right]}/{(\mu-\lambda)^2}$ 
and
\[
\delta_2= -\frac{1}{\mu^3(1-\rho)} \E\left( \sum_{i=1}^{N} \sum_{k=1}^{N^{\prime}}   p (X(0)) p (X(\wT  -D_i + D_k^{\prime} ))\right ),
\]
where, as in ~\eqref{a-}, $(N,D_{1},\ldots,D_{N})$ and $(N',D_{1}',\ldots,D_{N'}')$ denote the number of
departures and the departure   times in the busy periods of length $\wT$ and $\wB_{1}$,
respectively.  The terms $\delta_1$ and $\delta_2$ are non-positive. Thus, at the first
order, the mean of $\pT$ is larger than the mean of $\wT$.
The following proposition which readily follows, compares the length of the busy-period of
the P-Queue with the mean of the length of the busy-period $\hT$
in an $M/M/1$ queue with  service rate $\mu+\eps \E_\nu[p(X(0))]$.
\begin{prop}[Comparison with reduced service rate]\label{cheval}
If $\hT$ is the length of a busy period of an $M/M/1$ queue with service rate
$\mu+\eps \E_\nu[p(X(0))]$ then
\begin{equation}\label{Lecheneaut}
\lim_{\eps\to 0}\frac{1}{\eps^2}\E\left(\hT-\pT \right)  =-\frac{1}{\mu^3(1-\rho)}\E\left(
\sum_{i=1}^{N} \sum_{k=1}^{N^{\prime}}  C_p\left( X(\wT  -D_i + D_k^{\prime})\right)\right ) ,
\end{equation}
where, as in Equation~\eqref{a-}, $(N,D_{1},\ldots,D_{N})$ and $(N',D_{1}',\ldots,D_{N'}')$ denote the number of
departures and the departure   times in the busy periods of length $\wT$ and $\wB_{1}$,
respectively 
and for $u\geq 0$,
$$C_p(u)=\E_\nu\left[p(X(0))p(X(u))\right]-\E_\nu\left[p(X(0))\right]^2$$ is, up to the
factor $\eps^2$, the covariance function of the capacity of the perturbed queue. 
\end{prop}

This result implies that, as for a non-negative
 perturbation function, 
 the variation of the service rate
has a negative impact on the  performance of the system.  The following result holds.

\begin{prop}[Negative impact of the variation of the service rate]
When the environment is positively correlated (when the function $u \rightarrow
C_p(u)$ is non-negative), then the first term of the expansion of $\E\left(\hT-\pT
\right)$ in $\eps$ is of order 2 and  negative.
\end{prop}
Comparing to the case of a non-negative perturbation function,
if the correlation function of the
environment is exponentially decreasing, a simple close expression for 
the right hand side member of Equation~\eqref{Lecheneaut}  seems to be difficult to
obtain, though the same qualitative results hold. 
\begin{prop}[Exponential decay]
When the correlation function of the
environment is exponentially decreasing, i.e. when $$C_p(x)=\Var[p(X(0))]\,e^{-\alpha x},
\qquad x\geq 0,$$ and some $\alpha >0$,  the function 
\[
\alpha\to \lim_{\eps\to 0} \frac{1}{\eps^2}\,\E\left(\hT-\pT \right)
\] is non-positive, non-decreasing and
concave. Moreover when $\alpha$ tends to infinity, this quantity converges to zero.
\end{prop}

\subsection{Fast Environments}
A general perturbation function $p$ is considered together with some stationary Markov
process $(X(t))$ with invariant probability distribution $\nu$. It is assumed that it
verifies a mixing condition such as
\begin{equation}\label{chab}
\lim_{t\to+\infty} |\E_\nu[f(X(0))g(X(t))]-\E_\nu[f(X(0))]\E_\nu[g(X(0))]|=0,
\end{equation}
for any Borelian bounded functions $f$ and $g$ on the state space ${\cal S}$. Note that
this condition is not restrictive in general since it is true for any ergodic Markov process
with a countable (or finite) state space or for any ergodic diffusion on $\R^d$, $d\geq
1$.

In this section, the environment is accelerated by a factor $\alpha>0$, described by the process  $(X(\alpha t))$. The behavior when $\alpha$ goes to infinity is
investigated. Note that when $\alpha$ goes to $0$, the environment is frozen: the service
rate remains constant and equal to $\mu+\eps p(X(0))$. Such a situation has also been
analyzed  by  Delcoigne \etal~\cite{Proutiere} through stochastic bounds. 

At the intuitive level, when $\alpha$ gets large, for $t$ and $h>0$ the total service capacity available
during $t$ and $t+h$ is given by
\[
\mu h +\eps \int_{t}^{t+h} p(X(\alpha u))\,du \stackrel{\text{dist.}}{=}
\mu h +\eps \frac{1}{\alpha}\int_{0}^{\alpha h} p(X(u))\,du\sim (\mu+\eps\E_\nu(p(X(0))))h
\]
by the ergodic Theorem. Thus, speeding up the environment averages the capacity of the
variable queue. This intuitive picture is rigorously established in the following proposition. 
\begin{prop}
When the environment is given by $(X(\alpha t))$ and Equation~\eqref{chab} holds then if
$\delta_2(\alpha$) is the $\eps^2$ coefficient of the expansion of  with respect to $\eps$,
\[
\E(\pT-\wT )= \frac{\E_\nu\left(p(X(0))\right)}{(\mu-\lambda)^2}\,\eps +\delta_2(\alpha)\,\eps^2 +o(\eps^2),
\]
the following  relation holds,
\[
\lim_{\alpha\to+\infty} \delta_2(\alpha)= \frac{\E_\nu\left(p(X(0))\right)^2}{(\mu-\lambda)^3}.
\]
\end{prop}
\begin{proof}
The quantity $\delta_2(\alpha)$ is equal to $a_--a_+$ where $ a_-$ and $a_+$ are given by
Equations~\eqref{a-} and \eqref{a+}, respectively. We shall deal only with the first term
of $a_-$ in Equation~\eqref{a-}. Let
\[
F(\alpha)\stackrel{\text{def.}}{=} -\E \left ( \sum_{i=1}^N   \int_0^{\wT  + \wB_1} p^- (X(\alpha D_i))  p^+
   (X(\alpha s))  \, ds
   \right  )
\]
where $N$ is the number of customers in the busy period of length $\wT$ and   their 
departure times are denoted by $(D_i, 1\leq i\leq N)$. We have
\[
F(\alpha) = -\E \left ( \sum_{i=1}^N   \int_0^{\wT  + \wB_1} 
\E\left(p^- (X(\alpha D_i))  p^+ (X(\alpha s)) \mid \wT, N\right) \, ds
   \right  ).
\]
Relation~\eqref{chab} and the boundedness of $p$ (Assumption~($\mathrm{H_1}$)) show that,
almost surely, 
\[
\lim_{\alpha \to +\infty} \E\left(p^- (X(\alpha D_i))  p^+ (X(\alpha s)) \mid \wT,
  N\right) \, =\E_\nu\left(p^-(X(0))\right)\,\E_\nu\left(p^+(X(0))\right),
\]
therefore Lebesgue's theorem gives
\begin{align*}
\lim_{\alpha \to +\infty} &\frac{-F(\alpha)}{\E_\nu(p^-(X(0)))\,\E_\nu(p^+(X(0))) }  = 
\E\left(N\wT\right)+\E\left(\wB_1\right)\E\left(N\right) \\
&= \frac{1+\rho}{\mu(1-\rho)^3}+\frac{1}{\mu-\lambda} \frac{1}{1-\rho}
= \frac{2}{\mu(1-\rho)^3},
\end{align*}
by using the expressions of $\E(N)$ and $\E(N\wT)$  in the Appendix.
Similar calculations can be conducted for all the other terms to finally give  the Proposition. 
\end{proof}

\section{Appendix: Some useful  quantities for the $M/M/1$ queue}
Let $(A_k)$ (resp. $(D_k)$) denotes the arrival times (resp. departure times) in a busy
period of an $M/M/1$ queue with arrival rate $\lambda$ and service rate $\mu$. A busy
period denoted by $\wT$ that starts at time 0 will last a time $t$ and will consist of $N$
services 
if, and only if,
\begin{itemize}
\item[(i)] there are $(N-1)$ arrivals in $(0,t)$;
\item[(ii)] $D_N=t$;
\item[(iii)] $A_{k+1} \leq D_k$, $k=1,\ldots,N-1$.
\end{itemize}
If conditions $(i)$ and $(ii)$ are satisfied then $(A_2,\ldots,A_{N})$ and $(D_1,\ldots,D_{N-1})$
are independent and represent the ordered values of two sets of $N-1$ uniform $(0,t)$ 
random variables. Hence,
\begin{multline*}
b_n(t)=d\P(\wT<t,N=n)/dt = \frac{e^{-\lambda t}(\lambda t)^{(n-1)}}{(n-1)!}
\frac{\mu e^{-\mu t}(\mu t)^{(n-1)}}{(n-1)!}\\
\times \P(A_2 \leq D_1, \ldots ,  A_{n} < D_{n-1}).
\end{multline*}

The first two moments of the stationary busy period are given by
\[
\E(\wB_1) =\frac{1}{\mu-\lambda},\qquad  \E(\wB_1^2) =\frac{2}{\mu^2(1-\rho)^3}.
\]
Expression~(2.40) p.190 of Cohen~\cite{Cohen:01} shows that
\[
\varphi(z,\xi)=\sum_{n=1}^{+\infty}z^n \int_0^{+\infty} e^{-\xi t}b_n(t)\,dt,
\]
%where %$B_n(t) =\P (\wT<t,N=n)$ and
% $b_n(t) =d\P(\wT<t,N=n)/dt$
is given by 
\[\varphi(z,\xi)=\frac{1}{2\rho}\left(1+\rho+\mu^{-1}\xi-\sqrt{(1+\rho+\mu^{-1}\xi)^2-4\rho z}\right)
\]
for $|z|\leq 1$, $\Re(\xi)\geq 0$. It is easy to derive
\begin{align*}
\E(N)=\int_0^{+\infty} dt \sum_{n=1}^{+\infty} nb_n(t) & = \frac{1}{1-\rho},\\
\E(N\wT)=\int_0^{+\infty} t dt \sum_{n=1}^{+\infty} nb_n(t) & = -\frac{d^2 \varphi}{dzd\xi}(1,0) =
\frac{1+\rho}{\mu(1-\rho)^3},\\
\E[N(N-1)]=\int_0^{+\infty} dt \sum_{n=1}^{+\infty} n(n-1)b_n(t) & = \frac{d^2 \varphi}{dz^2}(1,0)
= \frac{2\mu^2\lambda}{(\mu-\lambda)^3}.
\end{align*}

To conclude one has to compute $\E(D)$ where $D=D_1+D_2+\cdots+D_N$. By using the
classical branching argument for the busy-period of the $M/M/1$ queue (see
Robert~\cite{Robert:08} for example), one gets 
\[
D=\sigma +\sum_{i=1}^{N_{\sigma}} \left(
\left(\sigma+\sum_{j=1}^{i-1}\wT_j\right)N_i+D_i\right),
\]
where $\sigma$ is the service time of the first customer of the busy-period, $N_{\sigma}$
the number of arrivals in the interval $[0,\sigma]$, $\wT_i$ the busy-period generated by the
$i$th customer arrived during $\sigma$, $N_i$ the number of customers in $\wT_i$,
$D_i$ the sum of the departure times of $\wT_i$ from the beginning of this
busy-period. Taking the expectation, it is easy to derive that
\[
\E(D)=\E(\sigma)+\E(\sigma
N_{\sigma})+\E(B)\E(N_{\sigma}(N_{\sigma}-1)/2)\E(N)+\E(\sigma
N_{\sigma})\E(D),
\]
where $N_{\sigma}$ has a geometric distribution with parameter
${\lambda}/{(\lambda+\mu)}$. Thus $$\E(N_{\sigma}(N_{\sigma}-1))=2\rho^2.$$ Simple
algebra  gives $\E(D)=\mu^2/(\mu-\lambda)^3$.

\providecommand{\bysame}{\leavevmode\hbox to3em{\hrulefill}\thinspace}
\providecommand{\MR}{\relax\ifhmode\unskip\space\fi MR }
\providecommand{\MRhref}[2]{%
  \href{http://www.ams.org/mathscinet-getitem?mr=#1}{#2}
}
\providecommand{\href}[2]{#2}


\begin{thebibliography}{10}

\bibitem{Agrawal:01}
Rajeev Agrawal, Armand~M. Makowski, and Philippe Nain, \emph{On a reduced load
  equivalence for fluid queues under subexponentiality}, Queueing Systems.
  Theory and Applications \textbf{33} (1999), no.~1-3, 5--41.

\bibitem{Altman}
E.~Altman, K.~Avrachenkov, and R.~Núñez~Queija, \emph{Perturbation analysis for
  denumerable markov chains with application to queueing models}, Advances in
  Applied Probability \textbf{36} (2004), no.~3, 839--853.

\bibitem{Antunes:02}
Nelson Antunes, Christine Fricker, Fabrice Guillemin, and Philippe Robert,
  \emph{Integration of streaming services and tcp data transmission in the
  {I}nternet}, Performance'05 (Juan les Pins), IFP WG 7.3, 2005.

\bibitem{Boxma:07}
O.~J. Boxma and I.~A. Kurkova, \emph{The {$M/M/1$} queue in a heavy-tailed
  random environment}, Statistica Neerlandica. Journal of the Netherlands
  Society for Statistics and Operations Research \textbf{54} (2000), no.~2,
  221--236.

\bibitem{Cohen:01}
J.~W. Cohen, \emph{The single server queue}, 2nd ed., North-Holland, Amsterdam,
  1982.

\bibitem{Proutiere}
F.~Delcoigne, A.~Proutière, and G.~Régnié, \emph{Modeling integration of
  streaming and data traffic}, ITC specialist seminar on IP traffic (Würzburg,
  Germany), July 2002.

\bibitem{Fricker:10}
C.~Fricker, F.~Guillemin, and P.~Robert, \emph{Perturbation analysis of an
  {M/M/1} queue in a diffusion random environment}, preprint, January 2004.

\bibitem{Grandell:01}
J.~Grandell, \emph{Point processes and random measures}, Advances in Applied
  Probability \textbf{9} (1977), 502--526.

\bibitem{Predag}
Predag Jelenkovi\'c and Petar Mom\v{c}ilovi\'c, \emph{Resource sharing with
  subexponential distributions}, Infocom'2002 (New York), June 2002.

\bibitem{Massoulie}
L.~Massouli\'e and J.~Roberts, \emph{Bandwidth sharing: Objectives and
  algorithms}, {INFOCOM }'99. Eighteenth Annual Joint Conference of the IEEE
  Computer and Communications Societies, 1999, pp.~1395--1403.

\bibitem{Nunez3}
R.~Núñez-Queija, \emph{Sojourn times in a processor sharing queue with service
  interruptions}, Queueing Systems \textbf{34} (2000), 351--386.

\bibitem{Nunez2}
\bysame, \emph{Sojourn times in non-homogeneous {QBD} processes with processor
  sharing}, Stochastic Models (2001), 61--92.

\bibitem{Nunez1}
R.~Núñez-Queija and O.J. Boxma, \emph{Analysis of a multi-server queueing model
  of {ABR}}, J. Appl. Math. Stoch. An. \textbf{11} (1998), 339--354.

\bibitem{Robert:08}
Philippe Robert, \emph{Stochastic networks and queues}, Stochastic Modelling
  and Applied Probability Series, vol.~52, Springer, New-York, June 2003.

\end{thebibliography}
\end{document}